\documentclass[letterpaper]{article}
\usepackage{soul}
\soulregister{\cite}{7}
\usepackage[T1]{fontenc}

\usepackage{geometry}
\usepackage{setspace}
\renewcommand{\st}[1]{}

\usepackage[style = chem-acs, articletitle = true, chaptertitle = true, doi = true]{biblatex}
\usepackage{graphicx}
\usepackage{float}
\newfloat{scheme}{htbp}{los}
\floatname{scheme}{Scheme}
\floatname{chart}{Chart}
\newfloat{graph}{htbp}{loh}

\usepackage{chemformula} 
\usepackage[version = 4]{mhchem} 

\usepackage{authblk}
\author[ a]{Ali Najjar Amiri}
\author[ b]{Trevor Kling}
\author[ c]{David Barton}
\author[ a,b\textsuperscript{*}]{Mahdi Hosseini}
\affil[a ]{Department of Electrical and Computer Engineering, Northwestern University, Evanston, IL, 60208, U.S.A}
\affil[b ]{Applied Physics Program, Northwestern University,  Evanston, IL, 60208, U.S.A}
\affil[c ]{Department of Materials Science and Engineering, Northwestern University,  Evanston, IL, 60208, U.S.A}
\title{Collective Radiative Enhancement of Rare-Earth Ions in Lithium Niobate via Engineered Large-Area Nanohole Arrays}
\date{*Email: mh@northwestern.edu}

\begin{document}

\maketitle

\begin{abstract}
\st{Enhancing light–matter interactions is a central goal in developing low-loss classical and quantum photonics.} Conventional approaches \textcolor{black}{to light-matter interactions} rely on \st{optical cavities or meta-material nanostructures to} \textcolor{black}{engineering}  photonic density of states. More recently, tailoring the spatial geometry of atoms or emitters themselves has emerged as a powerful and complementary route to control collective radiative properties. Here we experimentally realize a geometry-engineered ensemble of rare-earth ions by fabricating a periodic array of subwavelength gold nanoholes on lithium niobate implanted with thulium ions, forming a semi–two-dimensional array of quantum emitters embedded in a high-index crystalline thin film. The hybrid structure can simultaneously support localized and lattice plasmon resonances from the metallic array and collective atomic resonances from the ion ensemble. Using time-resolved photoluminescence and temperature-dependent measurements, we observe enhanced radiative emission attributed to collective atomic effects mediated by the nanohole lattice, distinct from single-emitter Purcell enhancement. Our results demonstrate a new regime of light–matter interaction \st{where plasmonic, lattice, and collective atomic resonances coexist and interact, } opening a pathway toward broadband and scalable, geometry-controlled quantum optical interfaces in solid-state platforms.
\end{abstract}


The ability to engineer and enhance light–matter interactions is critical for a wide range of quantum technologies, including quantum information processing \cite{welte2018photon}, quantum memories \cite{lei2023quantum}, single-photon sources \cite{esmann2024solid}, precision sensing \cite{bohr2024collectively}, transduction \cite{xie2025scalable} and distributed quantum networks \cite{knaut2024entanglement, ourari2023indistinguishable, ruskuc2025multiplexed}. Traditionally, such enhancement has been achieved by modifying the photonic environment of emitters, most notably through high-finesse optical cavities or metamaterial nanostructures that reshape the local density of optical states. While these approaches have enabled remarkable progress, they often rely on strong field confinement at the expense of scalability, bandwidth, or coherence, particularly in solid-state systems.
An emerging and conceptually distinct strategy focuses instead on engineering the spatial geometry of emitters themselves \cite{ pak2022long, shahmoon2017cooperative, chang2012cavity, zhou2024trapped, srakaew2023subwavelength}. When emitters are arranged in extended or structured ensembles with inter-emitter separations comparable to the optical wavelength, collective radiative effects arise that have no counterpart in single-emitter systems. The collective effects sought include, but not limited to, superradiant and subradiant states, cooperative frequency shifts, long-range photon-mediated interactions, and directional emission. Such collective phenomena are fundamentally many-body in nature and offer new opportunities for controlling light–matter interactions beyond what is achievable with cavities or isolated nanoantennas. Recent theoretical work has shown that extended and low-dimensional atomic arrays can exhibit highly tunable optical responses, including narrow cooperative resonances and suppressed spontaneous emission \cite{masson2024dicke, masson2022universality, asenjo2017exponential, ruostekoski2023cooperative, rui2020subradiant, kling2025cooperative, kling2023characteristics, kundu2025cooperative}, governed primarily by geometry rather than local field confinement.
In parallel, plasmonic nanostructures and periodic metallic lattices have been extensively studied as platforms for manipulating light on subwavelength scales \cite{guan2022light, garcia2007colloquium}. Arrays of metallic nanoparticles or nanoholes support not only localized surface plasmon resonances, but also collective surface lattice resonances (SLRs) arising from diffractive coupling within the array. These lattice modes can exhibit high quality factors, strong field enhancement, and long in-plane propagation lengths (controlled by the lattice size), especially when embedded in high-index or asymmetric dielectric environments \cite{bin2021ultra,utyushev2021collective, yang2019narrow}. The coexistence of localized plasmon modes and lattice resonances produces a rich optical landscape characterized by hybridization \cite{prodan2003hybridization,murray2004transition}, Fano interference \cite{luk2010fano}, and strong sensitivity \cite{auguie2008collective} to geometry and refractive index.
Despite extensive studies of lattice and collective atomic resonance effects, their interplay remains largely unexplored, particularly in solid-state platforms hosting dense ensembles of quantum emitters. A central open question is whether geometry-engineered emitter ensembles can access new regimes of cooperative light–matter interaction without relying on conventional cavities. It is also important to study how collective atomic resonances interact with other photonic and lattice modes when both are present.
In this work, we experimentally study such behavior by realizing a hybrid system consisting of a periodic array of gold nanoholes fabricated on lithium niobate (LN)  implanted with thulium (Tm) ions used as quantum centers. The nanohole array serves a dual purpose: it supports localized and lattice plasmonic resonances, and it acts as an implantation mask that creates a spatially modulated, semi–two-dimensional array of rare-earth ions beneath the holes. \textcolor{black}{The gold layer can be subsequently removed after implantation to reduce the resonant modes to only the atomic-lattice mode.} LN provides a high-index, low-loss crystalline host with excellent optical and electro-optic properties, while Tm ions offer long-lived optical transitions suitable for quantum applications \cite{pak2022long,dutta2023atomic, labbe2025thin}.
By combining time-resolved photoluminescence (PL) measurements with temperature-dependent studies, we observe radiative enhancement that cannot be explained solely by single-emitter Purcell effects or plasmonic nonradiative decay. Instead, our results are consistent with the emergence of collective atomic resonances, whose strength and manifestation depend on the emitter geometry, density, and coherence, and \textcolor{black}{can be} mediated by the plasmonic and lattice modes of the nanohole array. The observed behavior highlights a regime in which plasmonic structures do not merely enhance local fields, but act as a scaffold that enables and shapes collective emission from an extended atomic ensemble. \textcolor{black}{We observe that, in the absence of the plasmonic resonance (i.e., after removal of the gold layer), the radiative enhancement is partially recovered due to the presence of the atomic-lattice resonance.}
Our findings establish a new hybrid platform where plasmonic, lattice, and collective atomic resonances coexist and interact, revealing rich physics at the intersection of nanophotonics and many-body quantum optics.

The experiment consists of a semi-2D array of Tm$^{3+}$ ions implanted into a thin-film lithium-niobate-on-insulator (LNOI) wafer. A thin gold mask is used to stop ions at certain locations, thus allowing implementation of an arbitrary 2D geometry. In practice, the Tm-Au reaction and finite mask feature size limit the purity of the geometry. The hybrid platform and the optical measurement configuration are illustrated in Figure~1. As shown in Figure~1, the gold nanohole array serves a dual function. First, it acts as an implantation mask that produces a spatially modulated, semi-2D distribution of implanted Tm$^{3+}$ ions beneath the nanoholes. The inset shows Stopping and Range of Ions in Matter (SRIM) simulations of the ion implantation profile for both unmasked (nanohole) and masked (metal-covered) regions of LNOI at an implantation energy of 100~keV. More details about the fabrication and implantation are provided in the Supplementary Note 1. In the unmasked regions, the ion distribution peaks at approximately $30$~nm below the LNOI surface, forming a thin, laterally structured layer of Tm ions that constitutes an effective atomic lattice. Second, the perforated metal layer can modify the light-atom interaction by supporting surface and lattice-mediated resonances. The relevant energy-level structure of Tm$^{3+}$ in LN is also shown in the inset of Figure~1. Here, the atoms are optically excited at a wavelength $\lambda = $794.62~nm, and emission is detected from the $^3\mathrm{H}_4 \rightarrow {}^3\mathrm{H}_6$ transition. Following excitation, ions can relax through multiple radiative pathways, including transitions to the $^3\mathrm{H}_5$, $^3\mathrm{F}_4$, and $^3\mathrm{H}_6$ manifolds. The branching ratio $\beta$ associated with the $^3\mathrm{H}_4 \rightarrow {}^3\mathrm{H}_6$ transition investigated in this work corresponds to a decay probability of approximately 0.73 \cite{sun2012optical, thiel2010optical}.

\begin{figure}[!t]
  \includegraphics{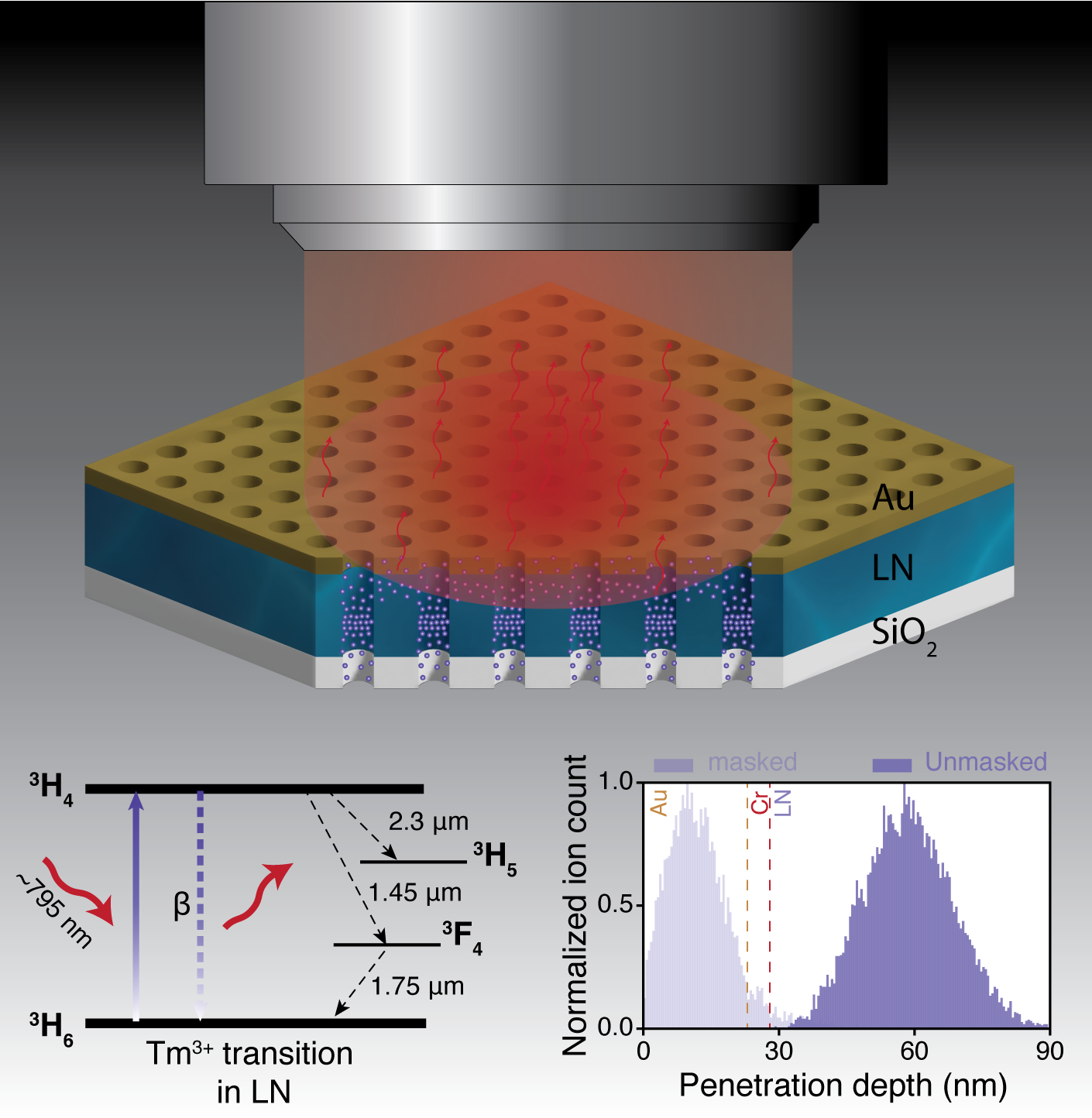}
  \centering
  \caption{\textbf{Schematic of thulium-implanted lithium niobate (LN) and the optical measurement setup.}
    Three-dimensional schematic of an engineered array of Tm$^{3+}$ ions implanted into LN, the 600-nm-thick top layer of a lithium-niobate-on-insulator (LNOI) wafer. Ion implantation is performed through a fabricated metal nanohole layer with a thickness of 28~nm. The bottom-left inset illustrates the relevant Tm$^{3+}$ energy levels in LN, where ions are excited at $\sim$795~nm and emission is detected from the $^3\mathrm{H}_4 \rightarrow {}^3\mathrm{H}_6$ optical transition. The bottom-right inset shows the simulated depth distribution of implanted Tm ions within the LN layer and the metal mask, calculated using the Stopping and Range of Ions in Matter (SRIM) software. For an implantation energy of 100~keV, the peak ion concentration is located approximately $30$~nm below the LN surface for the unmasked region (nanoholes), while for the masked region just $3.92\%$ of ions penetrate into LN. }
    
  \label{fgr1}
\end{figure}

Two complementary measurements are performed: (1) narrow-band excitation near the 794.62~nm transition for time-resolved PL measurements, with the emitted photons detected using a single-photon detector; and (2) broadband illumination using an unseeded tapered amplifier source, whose spontaneous emission spectrum spans 770--805~nm, to measure reflection spectra and probe pitch-dependent optical features of the nanohole arrays. Further details of the PL and reflection measurement procedures are provided in the Supplementary Note 2.

A central aspect of our experiment is that both the collective atomic response and the photonic lattice response (captured by the Rayleigh anomaly condition) depend on the same geometric parameter—the lattice pitch $a$. This shared dependence motivates the systematic investigation of pitch-dependent reflection spectra and PL lifetime dynamics presented in the following sections.
A spatially ordered array of emitters can be realized in solid-state platforms using focused ion implantation \cite{pak2022long} or high-resolution dry etching of nanostructures as implantation masks \cite{scarabelli2016nanoscale}. While these approaches offer precise spatial control, they are inherently time-consuming and limited in scalability, particularly for large-area structures. Focused ion implantation requires point-by-point writing with stringent alignment constraints, whereas nanohole dry etching introduces additional complexity related to sidewall roughness, etch selectivity, and damage to the host material. In contrast, the fabrication approach presented here combines parallel nanolithography with broad-beam ion implantation, providing a scalable and reproducible pathway to engineer geometry-defined, large-area emitter ensembles and broadband light-matter interactions in solid-state systems.

\begin{figure}[!t]
  \centering
  \includegraphics{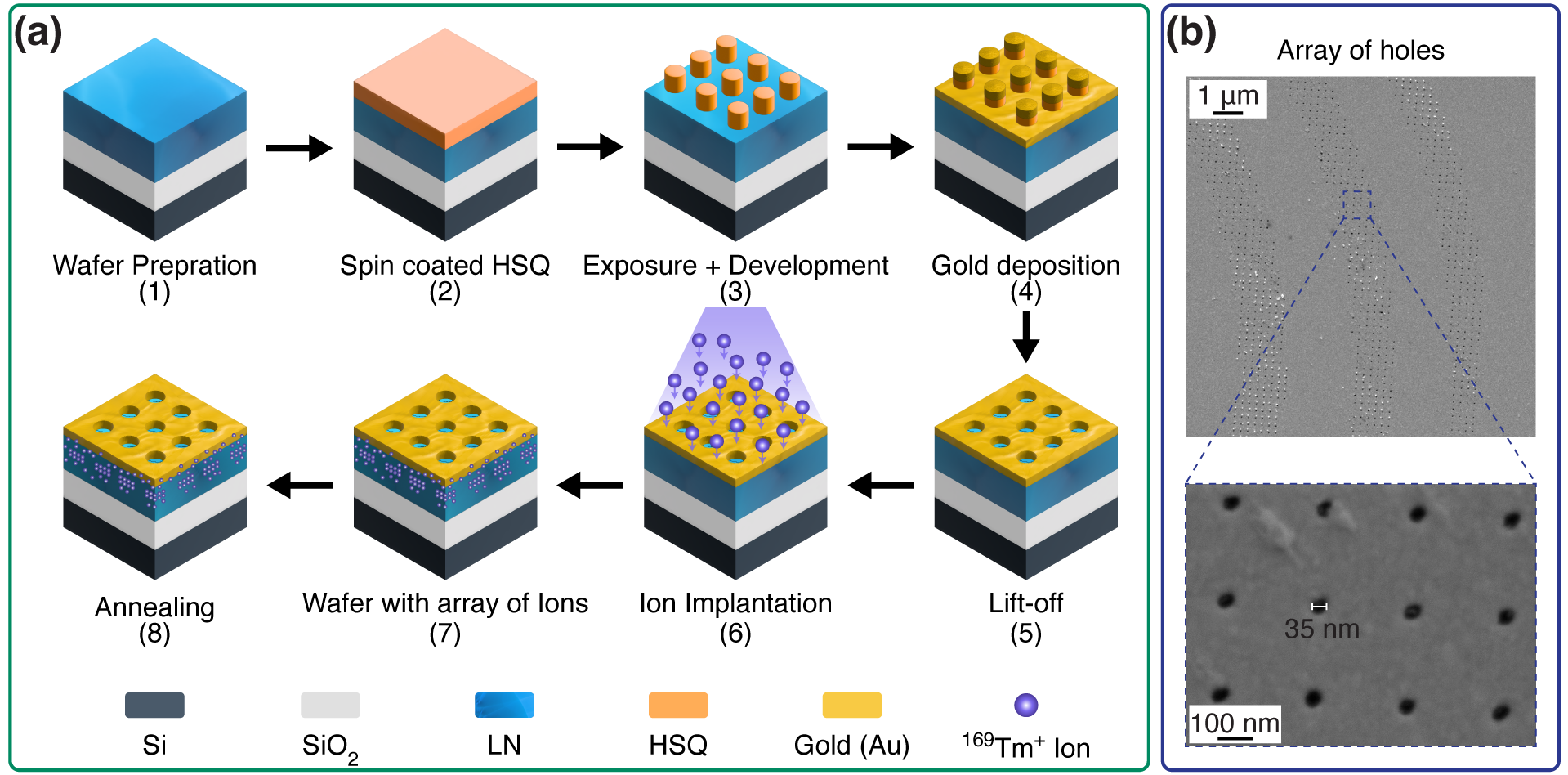}
\caption{\textbf{Nanofabrication process flow and SEM image of a metal nanohole array.}
\textbf{(a)} Schematic illustration of the nanofabrication process. 
(1) An LNOI wafer is prepared by standard cleaning followed by the application of an adhesion promoter. 
(2) A 115-nm-thick hydrogen silsesquioxane (HSQ) resist layer is spin-coated. 
(3) Electron-beam lithography (EBL) is used to pattern the resist, followed by development in a salty developer (1\% NaOH + 4\% NaCl) to form HSQ nanopillars. 
(4) A metal stack consisting of 5~nm Cr and 23~nm Au is deposited using electron-beam evaporation. 
(5) Lift-off is performed using buffered oxide etchant (BOE) 10:1 to define the metal nanohole array. 
(6) Broad-beam implantation of $^{169}$Tm$^{+}$ ions is carried out at an energy of 100~keV. 
(7) The ion-implanted wafer after implantation. 
(8) Post-implantation annealing at 600~$^\circ$C converts implanted ions into optically active $^{169}$Tm$^{3+}$. 
\textbf{(b)} Representative SEM image of a fabricated nanohole array after metal lift-off.
}
  \label{fgr2}
\end{figure}

Figure~2 summarizes the nanofabrication process used to realize the large-area nanohole arrays and the subsequent ion-implantation steps; see Supplementary Note~1 for full details. Based on SRIM simulations and the desired ion penetration depth, a total metal thickness of 28~nm was selected for the implantation mask. A $\sim$115~nm-thick HSQ resist layer was spin-coated onto the LNOI substrate, patterned by electron-beam lithography with systematically optimized pixel-dose exposures, and developed in an aqueous solution of 1\% NaOH and 4\% NaCl. A metal stack of 5~nm Cr and 23~nm Au was then deposited by electron-beam evaporation, and lift-off was performed using buffered oxide etchant (BOE) 10:1 with ultrasonic agitation and gentle mechanical cleaning to define the nanohole arrays \cite{wang2015high, duan2011direct}. The patterned samples were subjected to broad-beam ion implantation, followed by post-implantation annealing at 600~$^\circ$C for 6~h to activate the implanted Tm$^{3+}$ ions, as discussed in Supplementary Note~3. \textcolor{black}{The gold and chromium layers can be subsequently removed using standard wet etchants (see Supplementary Note~8).} Figure~2(b) shows a representative SEM image after lift-off, where hole diameters of approximately 35~nm are obtained for a trial pattern with a pitch of $\sim$211~nm. Using this process, we fabricated large-area nanohole arrays (each consisting of 25\,600 holes, $160 \times 160$), with lattice pitches of $0.6\lambda$ ($\sim$211~nm), $0.8\lambda$ ($\sim$282~nm), and $1.2\lambda$ ($\sim$423~nm), where $\lambda$ is the wavelength of excitation in LN.

\begin{figure}[!t]
    \centering
  \includegraphics{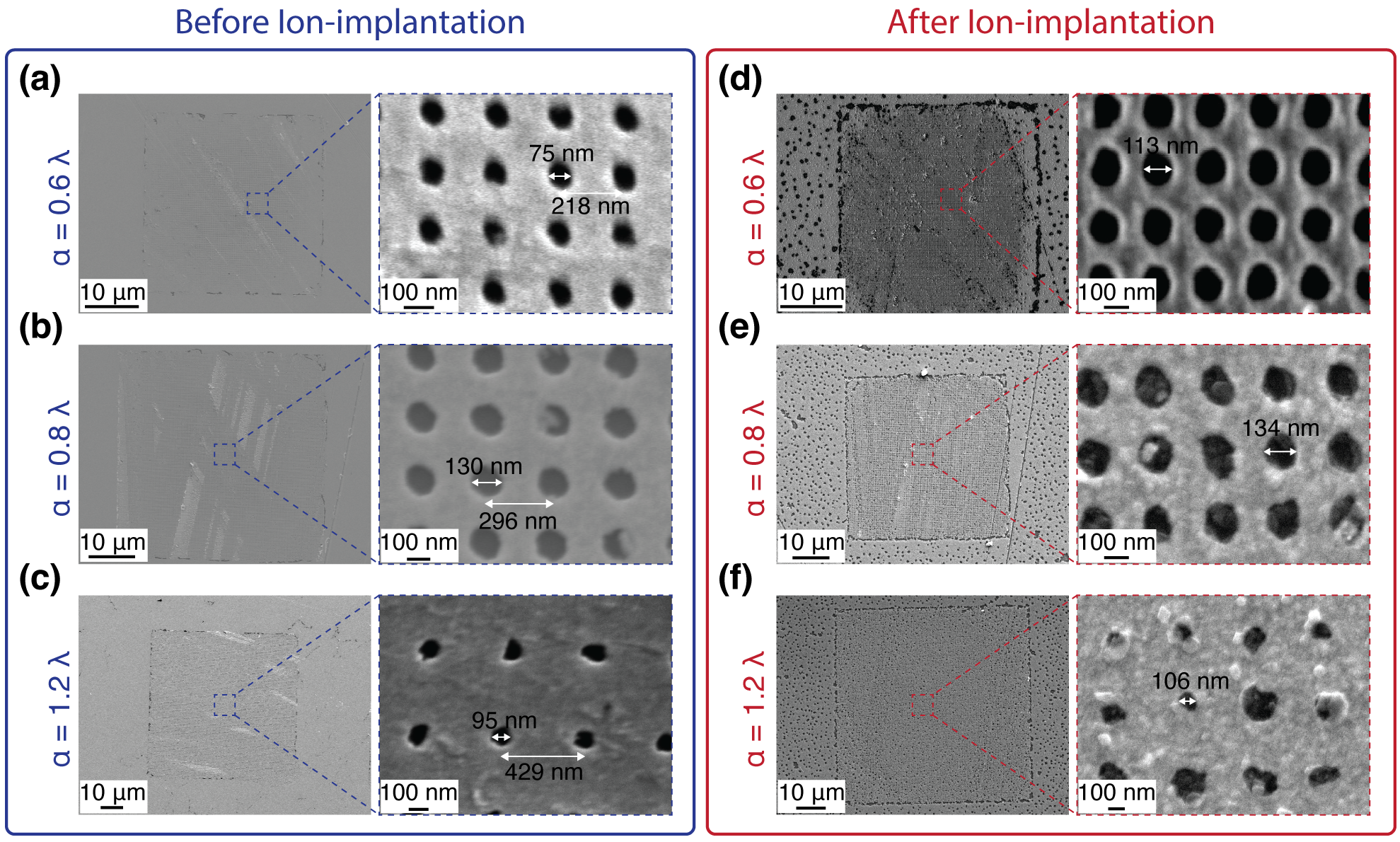}
\caption{\textbf{SEM images of nanohole arrays with different pitches before and after ion implantation.}
\textbf{(a--c)} SEM images of large-area nanohole arrays consisting of $160 \times 160$ holes with different lattice pitches. The measured hole diameters are 75~nm, 130~nm, and 95~nm for pitch values of $0.6 \lambda$, $0.8 \lambda$, and $1.2 \lambda$, respectively. 
\textbf{(d--f)} SEM images of the same arrays shown in (a--c) after ion implantation. The hole diameters increase due to ion-induced modification of the metal mask after annealing, with minimum hole diameters of 113~nm, 134~nm, and 106~nm for pitch values of $0.6 \lambda$, $0.8\lambda$, and $1.2 \lambda$, respectively. In addition, nonuniform holes with random sizes are generated primarily outside the patterned arrays, where continuous gold regions are present and are more strongly affected by the annealing process.
}
  \label{fgr3}
\end{figure}

Figure~3 presents SEM images of three large-area nanohole arrays before and after ion implantation. Figures~3(a)-(c) show representative SEM images of the as-fabricated arrays, together with the measured hole diameters and pitch values. For each pitch, a pixel-dose sweep in the range of 30-200~fC was performed, and the exposure dose yielding the highest pattern fidelity and hole yield across the full array area was selected. Based on this optimization, doses of 100~fC, 180~fC, and 200~fC were chosen for the lattice constant, $a$, of $0.6\lambda$, $0.8\lambda$, and $1.2\lambda$ , respectively. Although nanohole diameters as small as 35~nm can be achieved using the same fabrication process, as shown in Figure~2(b), the minimum hole diameters increase for large, densely packed arrays due to proximity effects during electron-beam exposure \cite{ren2004proximity}. The bright residual features observed in the SEM images arise from gold redeposition during the lift-off process or from partially intact gold sidewalls of the HSQ pillars. These residues were largely removed by gentle mechanical cleaning using IPA-soaked cotton swabs, as indicated by visible scrubbing marks in the SEM images. Importantly, many of the remaining residues do not fully obstruct the nanoholes, allowing ions to penetrate the substrate during implantation.

Following ion implantation and post-implantation annealing (Figures~3(d)-(f)), the minimum hole diameters increase further, and additional nonuniform holes emerge predominantly outside the patterned array regions. These morphological changes \textcolor{black}{can be} attributed to implantation- and annealing-induced restructuring of the thin gold film, as discussed in the Supplementary Note 3. The interaction between Au and implanted Tm ions is expected to be strongest in regions with excess gold, promoting enhanced atomic diffusion, grain-boundary migration, and partial dewetting during high-temperature annealing. Importantly, although the nanoholes appear larger after implantation and annealing, the effective ion-implantation region beneath each hole is expected to remain confined and closer in lateral extent to the original nanohole dimensions prior to implantation. This confinement arises because ion transmission is primarily determined by the open aperture of the hole during implantation, whereas the subsequent hole enlargement is a thermally driven surface effect occurring after implantation. As a result, the spatial distribution of implanted Tm ions preserves the intended lateral modulation set by the pre-implantation nanohole geometry, even as the metal morphology evolves during annealing. \textcolor{black}{Although the mask and nanoholes exhibit some spatial non-uniformity (particularly in the 1.2$\lambda$ case), we have verified that most of the surface morphology changes arise during the annealing process and following ion implantation. We therefore expect the implanted regions themselves to exhibit improved uniformity.} At the same time, annealing naturally produces adjacent gold regions with randomly distributed holes that lack long-range periodicity, providing disordered reference regions for comparison with the ordered nanohole arrays.

These disordered regions introduce non-radiative decay and loss but do not sustain coherent lattice sums or geometry-defined collective effects that rely on periodicity. We therefore compare the PL response of the ordered nanohole arrays with that of the adjacent disordered hole regions as reference to assess the role of lattice periodicity in shaping the observed lifetime dynamics. This contrast directly connects to the theoretical framework developed earlier \cite{kling2025cooperative} (see the Supplementary Note~3), where it was shown that ordered arrays enable coherent electromagnetic buildup through lattice sums ($S_{\mathrm{lat}}$) and phase-sensitive dipole-dipole interactions described by the Green’s-function formalism. In this framework, the collective atomic response can be expressed in terms of geometry-dependent frequency shifts and linewidth modifications,
\begin{equation}
\Delta_n + \frac{i}{2}\Gamma_n
=
-\frac{3\gamma\lambda_a}{2n_{\mathrm{ref}}}
\frac{N(\omega)}{2}
\sum_m
\mathbf{G}(k,\mathbf{r}_n,\mathbf{r}_m),
\tag{1}
\end{equation}
where $\Delta_n$ is the collective resonance shift, $\Gamma_n$ is the collective radiative decay modification, and $N(\omega)$ is the effective number of resonant emitters. Both quantities sensitively depend on the lattice spacing through the geometric sum of the Green’s function. In contrast, disordered features primarily introduce metal-induced non-radiative decay and radiation trapping without coherent interaction or well-defined momentum matching.

We selected 794.62~nm as the fixed excitation wavelength for all subsequent measurements, as it lies at (or very near) the peak of the Tm$^{3+}$ absorption profile at both room and cryogenic temperatures, ensuring near-optimal and consistent excitation of the ensemble (see Supplementary Note~4).

\begin{figure}[!t]
  \includegraphics[scale=0.98]{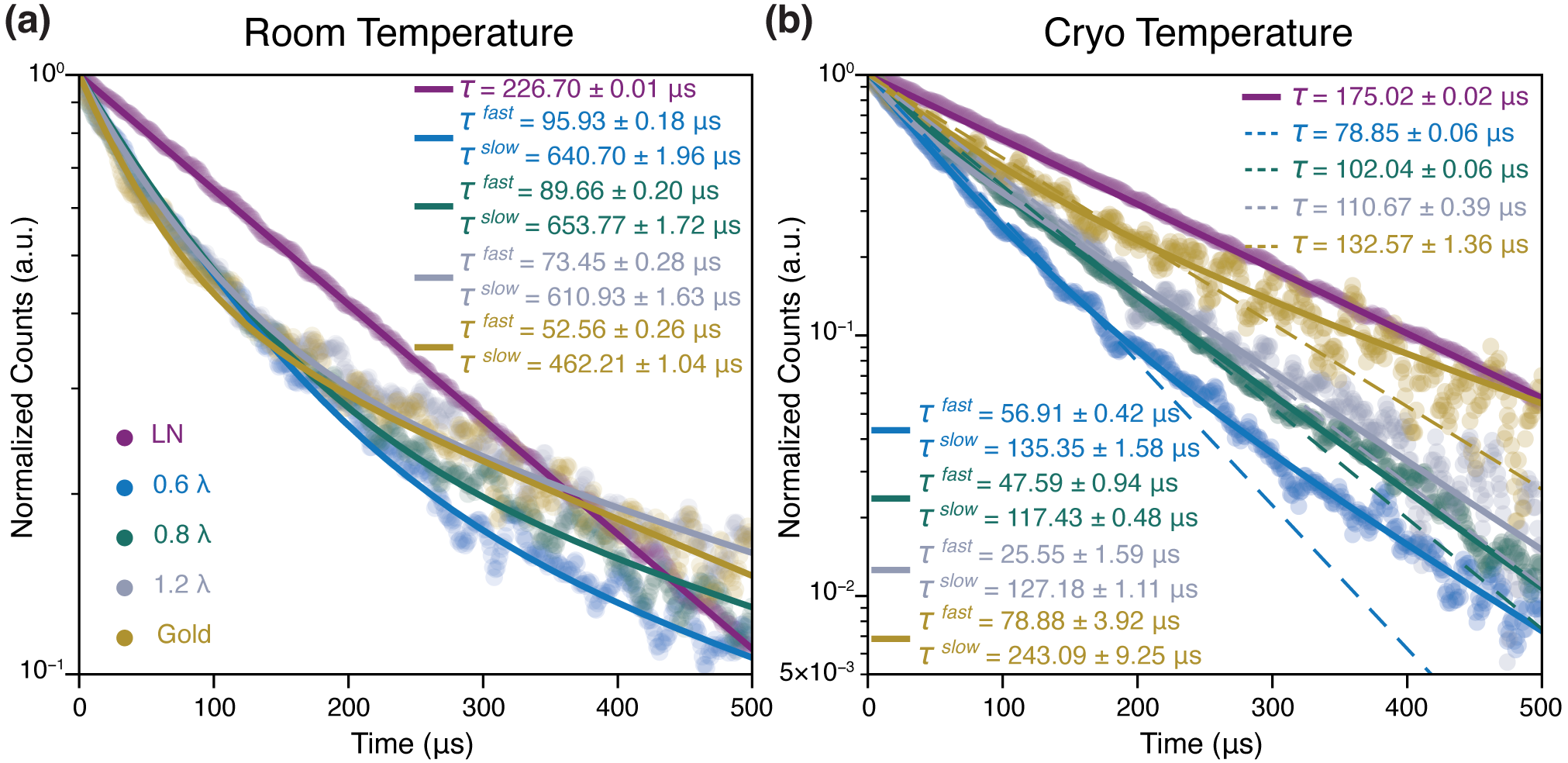}
\caption{\textcolor{black}{\textbf{Photoluminescence (PL) lifetime comparison between pure lithium niobate (LN), gold regions with disordered nanoholes, and ordered nanohole arrays.}
\textbf{(a)} Room-temperature PL decay traces. Pure LN exhibits a single-exponential decay ($A \exp\!\left(-t/\tau\right)$
), whereas regions containing ordered nanohole arrays display a bi-exponential decay ($A_{\mathrm{fast}} \exp\!\left(-t/\tau_{\mathrm{fast}}\right) +
A_{\mathrm{slow}} \exp\!\left(-t/\tau_{\mathrm{slow}}\right)$), indicating the coexistence of fast and slow decay channels associated with plasmonic\textcolor{black}{, nonradiative, and radiation trapping effects}. The fitted ratio $A_{\mathrm{fast}}/A_{\mathrm{slow}}$ is 1.44 for the disordered gold region, and 3.5, 2.7, and 1.8 for ordered arrays with pitch values of $0.6\lambda$, $0.8\lambda$, and $1.2\lambda$, respectively. At early decay times, ordered nanohole arrays with different pitch values show similar decay behavior; however, at longer times, the $0.6\lambda$ array exhibits the fastest decay, while the $1.2\lambda$ array shows the slowest decay. 
\textbf{(b)} Cryogenic PL decay traces. Although nonradiative decay processes are largely suppressed at cryogenic temperatures, the disordered gold regions continue to exhibit bi-exponential decay behavior, consistent with persistent plasmonic effects ($A_{\mathrm{fast}}/A_{\mathrm{slow}}$ = 1.34). The ordered nanohole arrays also require bi-exponential fitting, with $A_{\mathrm{fast}}/A_{\mathrm{slow}}$ ratios of 2.47, 0.32, and 0.31 for pitch values of $0.6\lambda$, $0.8\lambda$, and $1.2\lambda$, respectively. For relative comparison of the pitch-dependent trends, single-exponential fits (dashed lines) are applied to the ordered nanohole arrays and disordered gold decay curves. Based on these fits (with single time constant $\tau$), the extracted PL lifetimes indicate rate enhancements of 122\%, 72\%, and 58\% relative to pure LN, and rate enhancements of 68\%, 30\%, and 20\% relative to the disordered gold region, for pitch values of $0.6\lambda$, $0.8\lambda$, and $1.2\lambda$, respectively. The additional enhancement observed in the ordered arrays beyond the disordered reference highlights the role of lattice periodicity and photonic-atomic lattice hybrid modes in shaping the collective radiative dynamics.}
}

  \label{fgr4}
\end{figure}

Passive electromagnetic simulations and reflection measurements (see Supplementary Note 5) confirm the presence of photonic resonances in the nanohole arrays and establish a photonic baseline for the system. The modest Purcell enhancement and the weak pitch dependence of the absorption at 794.62~nm indicate that purely photonic effects are expected to play only a secondary role in modifying the emission dynamics. \textcolor{black}{Reflection measurements performed on a reference gold-on-LNOI sample without atoms (see Supplementary Note 5) reveal a photonic resonance near 780 nm. This resonance is slightly modified after annealing, likely due to surface deformation of the gold layer.} These considerations motivate the following time-resolved measurements, where we directly examine the PL lifetime and demonstrate that the dominant changes arise from collective atomic effects rather than from the passive photonic response. Time-resolved PL measurements reveal qualitatively different decay dynamics between uniformly ion-implanted LN and adjacent gold regions containing disordered nanoholes at both room and cryogenic temperatures. While pure LN exhibits a single-exponential decay, the gold regions show a robust bi-exponential behavior, with a fast component attributed to metal-induced nonradiative loss and a slower component persisting at low temperature, consistent with radiation trapping mediated by plasmonic modes at the gold–LN interface. Further details are provided in Supplementary Note 6. \textcolor{black}{While stretched exponential decay can arise from a broad distribution of decay rates \cite{lei2023many, solomon2024anomalous} in disordered or inhomogeneous ensembles, we find that the extracted stretching exponent $\beta \approx 1$ (see Supplementary Note 7), indicating that the decay is not governed by a wide continuum of lifetimes. Instead, the data are more naturally described by two dominant decay channels, consistent with the distinct electromagnetic environments created by the patterned metal layer. We therefore adopt a bi-exponential model, which provides a more direct physical interpretation of the fast and slow components observed in the measurements.}

\begin{figure}[!t]
  \centering
  \includegraphics{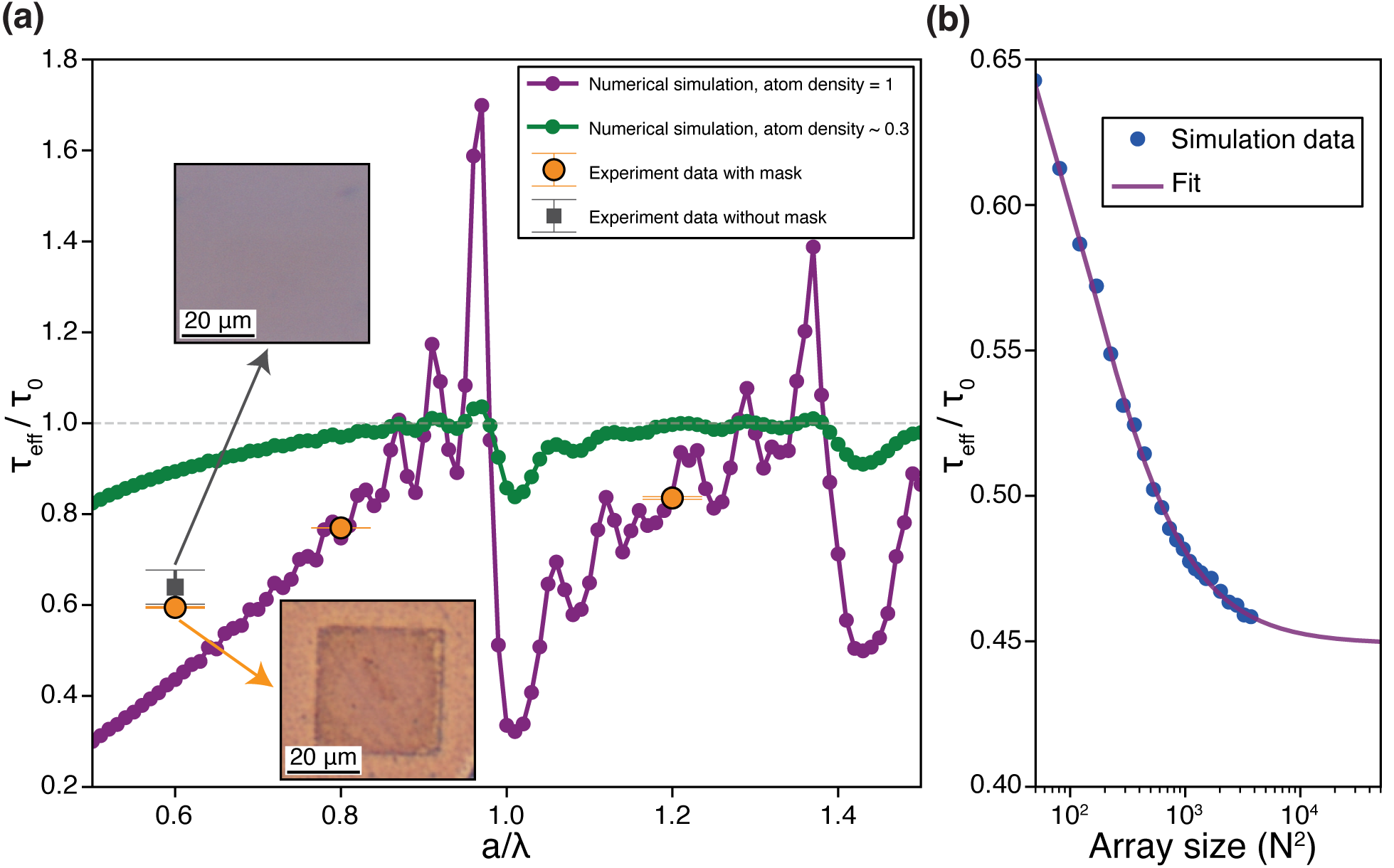}
\caption{\textcolor{black}{\textbf{Collective atomic lattice resonance.} (a) Numerical simulation result (obtained using numerical Green's function formalism) of lifetime modification together with experimental results are shown for different array spacing. Simulation considers $65 \times 65$ square lattice of Tm ions in lithium niobate (LN). Experimental results show good agreement with the theory. The inset optical images show the nanohole array before and after mask removal of corresponding data points. (b) Numerical simulation result of lifetime modification as a function of array size carried out using effective atomic density of 1 per site, considering $a/\lambda=0.6$. The fit function to the numerical data is $36.4/(a/\lambda+139.8)+0.45$.   } 
}
  \label{fgr5}
\end{figure}

Figure~4 presents time-resolved PL decay measurements from ordered nanohole arrays with lattice pitches of $0.6\lambda$, $0.8\lambda$, and $1.2\lambda$, compared with \textcolor{black}{both} pure ion-implanted LN \textcolor{black}{and adjacent gold regions containing disordered nanoholes}, at both room temperature and cryogenic temperature. In contrast to the single-exponential decay observed in pure LN, \textcolor{black}{both the disordered gold regions and} the ordered nanohole arrays exhibit a clear bi-exponential decay behavior at room temperature, indicating the coexistence of fast and slow decay channels. At room temperature (Figure~4(a)), the PL decay traces from the ordered arrays show bi-exponential dynamics, similar to those observed in the disordered gold regions discussed \textcolor{black}{in detail} in Supplementary Note 6. The extracted fast and slow lifetimes exhibit a weak dependence on the lattice pitch, as variations in pitch slightly modify the non-radiative decay channels and the strength of radiation trapping. No clear evidence of plasmonic resonant enhancement in the PL is observed, as any such effect (if exists) is masked by loss mechanisms in the metal at room temperature. We further note that atomic lattice effects are not expected to play a significant role under room-temperature conditions. At cryogenic temperatures (Figure~4(b)), the nonradiative decay channels and radiation trapping are strongly suppressed, resulting in a reduced degree of bi-exponential behavior. While a bi-exponential model is still required to accurately fit the full decay traces, the overall decay dynamics are less strongly influenced by metal-related loss mechanisms. This regime enables a more meaningful comparison of the effective radiative decay rates across different lattice pitches. For better comparison, single-exponential fits are applied to the cryogenic PL decay curves \textcolor{black}{of the ordered arrays and the disordered gold region}, focusing on the dominant decay channel. The extracted effective lifetimes \textcolor{black}{of the ordered arrays} show a clear dependence on lattice pitch, with the shortest lifetime observed for the $0.6\lambda$ array and progressively longer lifetimes for the $0.8\lambda$ and $1.2\lambda$ arrays. Relative to pure LN, these lifetimes correspond to PL rate enhancements of approximately 122\%, 72\%, and 58\% for the $0.6\lambda$, $0.8\lambda$, and $1.2\lambda$ arrays, respectively. \textcolor{black}{Importantly, the ordered arrays also exhibit rate enhancements of 68\%, 30\%, and 20\% relative to the disordered gold region for pitch values of $0.6\lambda$, $0.8\lambda$, and $1.2\lambda$ arrays, respectively. This additional enhancement beyond the disordered reference, which has undergone identical processing but lacks long-range periodicity, directly evidences the role of lattice-mediated collective interactions in modifying the radiative dynamics. To further confirm that the radiative enhancement can emerge solely from the atomic lattice, we removed the gold layer on one of the samples ($0.6 \lambda$) and compared PL data at cryo temperature (See Supplementary Note 8: Figure S7). As can be seen in Figure \ref{fgr5}(a), the radiative enhancement is mostly recovered even after removing the gold mask.}

This systematic trend with lattice spacing cannot be explained solely by the local plasmonic effects or radiation trapping, which are largely insensitive to long-range order. Instead, it is consistent with the emergence of collective atomic lattice resonances, most pronounced at low temperatures where coherence time is higher. In a periodic emitter ensemble, photon-mediated atom-atom coupling depends sensitively on lattice geometry through coherent Green’s-function summation, leading to geometry-dependent modifications of the collective radiative decay rate. These collective effects are most clearly revealed at cryogenic temperature, where dephasing and nonradiative losses are minimized. The photonic lattice can however help better guide the optical mode in the LN slab, thus indirectly enhancing the collective interactions \cite{kundu2025cooperative,boddeti2024reducing}.  Overall, the results in Figure~4 demonstrate that ordered nanohole arrays enable a regime in which plasmonic coupling, radiation trapping, and collective atomic interactions coexist, with lattice periodicity playing a decisive role in shaping the radiative dynamics. While room-temperature measurements are dominated by complex bi-exponential behavior arising from multiple competing mechanisms, cryogenic measurements isolate geometry-controlled radiative enhancement, providing direct evidence for lattice-mediated collective emission in this hybrid plasmonic-atomic system.

To understand the contribution of the collective atomic response to the observed decay curves, we performed simulations\cite{kling2025cooperative} of atomic interactions for a lattice of ions embedded in a substrate \textcolor{black}{(see Figure~5(a))}. With the exception of the total atom number, the parameters of the simulations are chosen to match those of the experimental devices based on SRIM simulations, nanohole diameters from SEM images, and spectral measurements of the materials \textcolor{black}{(See Supplementary Note 9)}.  \st{Figure~5 shows the resulting effective linewidth of the atoms, calculated as $\Gamma_{n} + \gamma$ for each atom $n$ and averaged over the ensemble to find $\Gamma_\mathrm{eff}$. Although the simulated $\Gamma_\mathrm{eff}$ is in accordance with the experimental values, the observed effect is an order of magnitude greater than the simulated effect due to the limited number of atoms used in the simulation.  The relative shape of the $\Gamma_\mathrm{eff}$ curve only depends on the choice of lattice geometry and implantation depth, as they contribute to the summation of the Green's functions in Eq.~1.}  

\textcolor{black}{Due to computational limitations, the lattice considered in these simulations is $65 \times 65$, smaller than the $160 \times 160$ site lattices considered in the experiments. As shown in Figure 5(b), increasing the lattice size beyond 65×65 does not significantly affect the results. We therefore limit our simulations to this smaller lattice to reduce the computational cost. The experimental data are in better agreement with the high-density-site model, in which an effective atom number of approximately one atom per homogeneous linewidth per site is assumed. We attribute the slight divergence of the $0.8\lambda$ and $1.2\lambda$ value from the exact simulated trend to a reduction in the spatial disorder compared to the simulated value for larger lattices, as these systems experience a reduced proximity effect during fabrication.}

\textcolor{black}{The lattice effects demonstrated here can be significantly increased through careful selection of the atomic environment.  LN demonstrates large inhomogeneous broadening, limiting the effective volume of atoms per site.  Isotopically-purified Silicon-28 offers a nanofabricatable material with exceptionally low inhomogeneous broadening \cite{berkman2025long}, which could improve the optically active atom density at each site and improve the lattice effect by several orders of magnitude.  Additionally, the use of a thin-film substrate with dielectric substrate thickness less than one wavelength can significantly extend the interaction range and enhance observed superradiant behavior \cite{kundu2025cooperative}.}
 
In summary, using thulium-implanted lithium niobate patterned with a periodic array of gold nanoholes, we observe modifications of the photoluminescence dynamics associated with the collective atomic resonances mediated by extended in-plane photonic modes. \st{The scalable engineering of atomic and photonic lattices offers new paradigms to study collective radiative interactions \cite{masson2022universality}, for applications in quantum state creation and  storage \cite{asenjo2017exponential}, and critical quantum sensing \cite{alushi2025collective}.}

\section{Acknowledgments}

We like to thank funding from National Science Foundation, Grant number: 2410054 and U.S. Department of Energy, Office of Science, Office of Advanced Scientific Computing Research, through the Quantum Internet to Accelerate Scientific Discovery Program under Field Work Proposal No. 3ERKJ381.
This work was carried out using the nanofabrication facilities of Northwestern University, including NUANCE and NUFAB, and was supported in part by NSF awards MRI (NSF DMR-1828676) and the SHyNE Resource (NSF ECCS-2025633), which provided access to the electron-beam lithography infrastructure. We also thank Serkan Butun and Shaoning Lu for insightful discussions and guidance related to the fabrication process.

\section{Supporting information}

Fabrication, optical measurements, model (Atomic lattice resonance, Photonic resonance, Radiation trapping, Atom implantation and surface interaction), optical excitation response as a function of
wavelength, photonic and plasmonic effects, disordered nanohole photoluminescence lifetime, extended photoluminescence lifetime and fit comparison, removing the gold mask to isolate the atomic-lattice effect, details of numerical simulations.


\printbibliography

\end{document}